\documentclass{ws-procs9x6}%
\usepackage{amsmath}%
\usepackage{amsfonts}%
\usepackage{amssymb}%
\usepackage{graphicx}
\providecommand{\U}[1]{\protect\rule{.1in}{.1in}}
\begin{document}

\title{A CPT-EVEN LORENTZ-VIOLATING NONMINIMAL COUPLING
BETWEEN FERMIONS AND PHOTONS}

\author{R.\ CASANA, M.M.\ FERREIRA JR.,$^*$
F.E.P.\ DOS SANTOS, E.O.\ SILVA}

\address{Departamento de F\'{\i}sica, Universidade Federal do Maranh\~{a}o\\
S\~{a}o Lu\'{\i}s, Maranh\~{a}o 65080-805,  Brazil\\
$^*$E-mail: manojr07@ibest.com.br}

\author{E.\ PASSOS\\}
\address{Departamento de F\'{\i}sica, Universidade Federal de Campina Grande\\
Caixa Postal 10071, 58109-970 Campina Grande, PB, Brazil\\}

\begin{abstract}
We propose a CPT-even and Lorentz-violating dimension-five nonminimal coupling
between fermionic and gauge fields, involving the CPT-even and
Lorentz-violating gauge tensor of the Standard-Model Extension. This
nonminimal coupling modifies the nonrelativistic regime of Dirac particles,
inducing new effects such as an electric-Zeeman-like spectrum splitting and an
anomalous-like contribution to the electron magnetic moment. These new effects
allow to constrain the magnitude of this nonminimal coupling in $1$ part in
$10^{16}.$

\end{abstract}

\bodymatter

\phantom{x}
\vskip10pt

\noindent
The Standard-Model Extension (SME) \cite{Colladay} is the natural framework
for studying properties of Lorentz violation (LV) in all sectors of the
Standard Model \cite{SME}. Some time ago, a Lorentz-violating and CPT-odd
nonminimal coupling between fermions and the gauge field was proposed
\cite{NM1} of the form $D_{\mu}=\partial_{\mu}+ieA_{\mu}+ig\epsilon
_{\mu\lambda\alpha\beta}(k_{AF})^{\lambda}F^{\alpha\beta}/2,$ in the context
of the Dirac equation, where $(k_{AF})^{\mu}=($v$_{0},\mathbf{v)}$ is the
Carroll-Field-Jackiw 4-vector and $g$ is the coupling constant. The
analysis of the nonrelativistic limit revealed that this coupling provides a
magnetic moment $\left(  g\mathbf{v}\right)  $ for uncharged particles
\cite{NM1}, and corrections to the hydrogen spectrum \cite{NMhydrog}.
Contributions to the dynamics of the Aharonov-Bohm-Casher problem
\cite{NMABC} and to the fermion-fermion ultrarelativistic scattering
\cite{NMmaluf} were also analyzed. In the present work, we propose a new
CPT-even \cite{NMscatt},\ dimension-five, nonminimal coupling linking the
fermionic and gauge fields in the context of the Dirac equation. The starting
point is a nonminimal coupling involving fundamental Dirac fermions and the
electromagnetic field in the context of the Dirac equation, $(i\gamma^{\mu
}D_{\mu}-m_{e})\Psi^{\left(  e\right)  }=0$, where $\Psi^{\left(  e\right)  }$
is the electron spinor wave function, and\textbf{\ }the covariant derivative
with nonminimal coupling is%
\begin{equation}
D_{\mu}=\partial_{\mu}+ieA_{\mu}+\frac{\lambda^{\left(  e\right)  }}{2}\left(
K_{F}\right)  _{\mu\nu\alpha\beta}\gamma^{\nu}F^{\alpha\beta},\label{covader}%
\end{equation}
where $\lambda^{\left(  e\right)  }$ is the electron nonminimal coupling
constant, and $\left(  K_{F}\right)  _{\mu\nu\alpha\beta}$ is the CPT-even LV
tensor of the photon sector of SME, which can be written in terms of four
$3\times3$ matrices $\kappa_{DE},\kappa_{DB},\kappa_{HE},\kappa_{HB}$:
$\left(  \kappa_{DE}\right)  _{jk}=-2\left(  K_{F}\right)  _{0j0k},$ $\left(
\kappa_{HB}\right)  _{jk}=\epsilon_{jpq}\epsilon_{klm}\left(  K_{F}\right)
_{pqlm}/2,$ $\left(  \kappa_{DB}\right)  _{jk}=-\left(  \kappa_{HE}\right)
_{kj}=\epsilon_{kpq}\left(  K_{F}\right)  _{0jpq}.$ The Dirac equation can be
explicitly written as
\begin{equation}
\left[  i\gamma^{\mu}\partial_{\mu}-e\gamma^{\mu}A_{\mu}+\frac{\lambda
^{\left(  e\right)  }}{2}\left(  K_{F}\right)  _{\mu\nu\alpha\beta}\sigma
^{\mu\nu}F^{\alpha\beta}-m_{e}\right]  \Psi^{\left(  e\right)  }%
=0,\label{DiracM1}%
\end{equation}
and $\sigma^{\mu\nu}=i(\gamma^{\mu}\gamma^{\nu}-\gamma^{\nu}\gamma^{\mu})/2.$
Thus, the relevant\ electron lagrangian is\textbf{\ }%
\begin{equation}
\mathrm{{\mathcal{L}}}_{(e)}=\bar{\Psi}^{(e)}(i{\rlap{\hbox{$\mskip 1 mu /$}}}%
\partial-e{\rlap{\hbox{$\mskip 1 mu /$}}}A-m_{e}+\frac{\lambda^{\left(
e\right)  }}{2}\left(  K_{F}\right)  _{\mu\nu\alpha\beta}\sigma^{\mu\nu
}F^{\alpha\beta})\Psi^{(e)},\label{Le}%
\end{equation}
with $\mathbf{\sigma}=(\sigma_{x},\sigma_{y},\sigma_{z})$ being the Pauli
matrices. In the momentum coordinates, $i\partial_{\mu}\rightarrow p_{\mu},$
the corresponding Dirac equation is
\begin{align}
i\partial_{t}\Psi^{(e)} &  =\left[  \mathbf{\alpha}\cdot(\mathbf{p}%
-e\mathbf{A)}+eA_{0}+m_{e}\gamma^{0}-\lambda^{\left(  e\right)  }i\gamma
^{j}\left(  \mathbb{E}^{j}+\mathbb{B}^{j}\right)  \right.  \nonumber\\
&  \left.  +\lambda^{\left(  e\right)  }\gamma^{0}\Sigma^{k}\left(
\mathbb{\tilde{E}}^{k}+\mathbb{\tilde{B}}^{k}\right)  \right]  \Psi
^{(e)}.\label{DiracM3}%
\end{align}
where we have used $F_{0j}=E^{j},F_{mn}=\epsilon_{mnp}B^{p}$, $\sigma
^{0j}=i\alpha^{j},$ $\sigma^{ij}=-\epsilon_{ijk}\Sigma^{k},$
\begin{equation}
\alpha^{i}=\left(
\begin{array}
[c]{cc}%
0 & \sigma^{i}\\
\sigma^{i} & 0
\end{array}
\right)  ,\text{ \ \ \ \ }\Sigma^{k}=\left(
\begin{array}
[c]{cc}%
\sigma^{k} & 0\\
0 & \sigma^{k}%
\end{array}
\right)  .
\end{equation}
and introduced the following definitions:%
\begin{eqnarray}
\mathbb{E}^{k}&=&( \kappa_{DE})_{kj}E^{j},
\qquad
\mathbb{B}^{k}=( \kappa_{DB})_{kj}B^{j},
\nonumber\\
\mathbb{\tilde {E}}^{k}&=&( \kappa_{HE})_{kq}E^{q},
\qquad
\mathbb{\tilde{B}}^{k}=( \kappa_{HB})_{kp}B^{p}.
\label{Edef1}%
\end{eqnarray}

In order to investigate the role played by this nonminimal coupling, we should
evaluate the nonrelativistic limit of the Dirac equation. At first order in
the Lorentz violating parameters, the following hamiltonian is achieved:
\begin{align}
H^{\left(  e\right)  }  &  =\frac{1}{2m_{e}}\left[  (\boldsymbol{p}%
-e\boldsymbol{A)}^{2}-e\left(  \boldsymbol{\sigma\cdot B}\right)  \right]
+eA_{0}+\lambda^{\left(  e\right)  }\boldsymbol{\sigma\cdot}\left(
\mathbb{\tilde{E}}+\mathbb{\tilde{B}}\right) \nonumber\\
&  -\frac{\lambda^{\left(  e\right)  }}{m_{e}}\left(  \mathbb{E+B}\right)
\cdot(\boldsymbol{\sigma}\times\mathbf{p})+\frac{e\lambda^{\left(  e\right)
}}{m_{e}}\left(  \mathbb{E+B}\right)  \mathbb{\cdot}(\mathbf{\sigma}%
\times\mathbf{A}), \label{HNR3}%
\end{align}
here written for uniform fields. This hamiltonian induces new effects to be
here discussed. The term $\mathbb{E}\cdot(\boldsymbol{\sigma}\times
\mathbf{p})$ is a generalization of the Rashba coupling term, $\mathbf{E}%
\cdot(\boldsymbol{\sigma}\times\mathbf{p}),$ while $\lambda^{(e)
}( \boldsymbol{\sigma\cdot}\mathbb{\tilde{B}})  $ implies a
straightforward tree-level contribution to the anomalous magnetic moment of
the electron, and the term $\lambda^{(e)}
( \boldsymbol{\sigma\cdot}\mathbb{\tilde{E}})$ leads to a kind of electric Zeeman effect. We begin discussing the
correction induced on the atomic spectrum of hydrogen by the term
$\boldsymbol{\sigma}\cdot\mathbb{\tilde{E}}$ involving the spin operator. In this case, the
correction energy is given by $\Delta_{E}=\lambda^{\left(  e\right)  }\langle
nljm_{j}m_{s}|\boldsymbol{\sigma\cdot}\mathbb{\tilde{E}}|nljm_{j}m_{s}%
\rangle.$ Now, we adopt a polarized spin configuration, $\boldsymbol{\sigma
}=\sigma_{z}\hat{z},$\ such that%
\begin{equation}
\Delta_{E}=\pm\lambda^{\left(  e\right)  }\left(  \kappa_{HE}\right)
_{3j}E_{j}\frac{m_{j}}{2l+1}, \label{EB}%
\end{equation}
with $E_{j}$\ being one of the components of the electric field, and $\left(
\kappa_{HE}\right)  _{3j}$\ a non-null element of the matrix $\left(
\kappa_{HE}\right)  $.\ The positive and negative signs correspond to
$j=l+1/2$\ and $j=l-1/2,$\ respectively. We also used $\langle
nljm_{j}m_{s}|\sigma_{z}|nljm_{j}m_{s}\rangle=m_{j}\hbar/(2l+1)$\textbf{,}
$\langle nljm_{j}m_{s}|\sigma_{x}|nljm_{j}m_{s}\rangle=\langle nljm_{j}%
m_{s}|\sigma_{y}|nljm_{j}m_{s}\rangle=0.$ The dependence on $m_{j}$ leads to a
spectrum splitting in $\left(  2j+1\right)  $ lines, representing an electric
Zeeman-like effect (due to the presence of an electric field, that can be
external or the atomic one). Regarding the possibility of measuring spectrum
shifts as small as $10^{-10}$ eV, and working with a typical atomic electric
field for the hydrogen fundamental level ($a_{0}\simeq0.529$ \AA ), whose
magnitude is $E\simeq5.1\times10^{11} N/C\simeq1.2\times10^{6} \left(
\text{eV}\right)  ^{2},$ the Zeeman-like splitting will be undetectable if
$\vert \lambda^{\left(  e\right)  }\left(  \kappa_{HE}\right)
_{3j}\vert E_{j}<10^{-10}$ eV. It leads to the
following upper bound:%
\begin{equation}
\big\vert \lambda^{\left(  e\right)  }\left(  \kappa_{HE}\right)
_{3j}\big\vert <8\times10^{-17}\left(  \text{eV}\right)  ^{-1}.
\label{bound1}%
\end{equation}

Another effect associated with the hamiltonian (\ref{HNR3}) is concerned with the
anomalous magnetic moment of the electron. The electron magnetic moment is
$\boldsymbol{\mu}=-\mu\boldsymbol{\sigma,}$ with $\mu=e/2m_{e}$, and $g=2$ the
gyromagnetic factor. The anomalous magnetic moment of the electron is given by
$g=2(1+a),$ with $a=\alpha/2\pi+...=0.00115965218279$ representing the
deviation \textbf{(}value in the year 2008\textbf{) }in relation to the usual
case. Hence, the magnetic interaction is $\boldsymbol{H}^{\prime}%
=\mu(1+a)\boldsymbol{\sigma}\cdot\mathbf{B}$. Very precise measurements and
QED calculations \cite{Gabrielse} yield $\Delta a\leq3\times10^{-11}.$ The
term $\lambda^{\left(  e\right)  }\boldsymbol{\sigma\cdot}\mathbb{\tilde{B}}$
provides tree-level LV\ contributions to the magnetic interaction, $\mu\left[
1+2m_{e}\lambda^{\left(  e\right)  }\left(  \kappa_{HB}\right)  _{33}%
/e\right]  \left(  \boldsymbol{\sigma}_{z}B_{0}\right)  $, for the magnetic
field along the z-axis, $\mathbf{B=}B_{0}\hat{z},$ and a spin-polarized
configuration in the $z$-axis. Such correction can not be larger than $a$, so
that
\begin{equation}
\big\vert \lambda^{\left(  e\right)  }\left(  \kappa_{HB}\right)
_{33}\big\vert \leq9.7\times10^{-11} \ ({\rm eV})^{-1},
\end{equation}
where we have used $m_{e}=5.11\times10^{5}$ eV, $e=\sqrt{1/137}.$ It is
important to mention that the bounds here found should not be confused with
the upper bounds on\textbf{ }the $\left(  K_{F}\right)  _{\mu\nu\alpha\beta}$
components already known in the literature, as in the present case the
constraint is on the magnitude of the CPT-even parameters as nonminimally
coupled. A complete investigation of the corrections on the hydrogen spectrum
implied by the hamiltonian (\ref{HNR3}) can be carefully carried out having as
counterpart the procedures known in the literature \cite{Hydrogen}. The
contributions implied by this nonminimal interaction to the photon self energy
were already evaluated \cite{Frede}. Also, the contributions to the fermion
self-energy (searching for possible modifications in the fermionic dispersion
relations) and to the vertex function are under analysis. These calculations
are relevant for establishing the 1-loop consistency of this model.

\section*{Acknowledgments}

The authors thank CAPES, CNPq, and FAPEMA for financial support.

\end{document}